\documentclass[11pt,twoside]{article}
\usepackage{asp2004}
\usepackage{psfig}
\usepackage{epsf}
\usepackage{graphics}
\usepackage{lscape}
\def\edcomment#1{\iffalse\marginpar{\raggedright\sl#1\/}\else\relax\fi}
\reversemarginpar

\begin{document}
\title{HI at a Redshift of Zero}
\author{Christopher M. Springob}
\affil{Center for Radiophysics and Space Research, Cornell University, 514 Space Sciences Building, Ithaca, NY 14853}

\begin{abstract}
While LOFAR and the SKA will enable the study of HI at the epoch of reionization for the first time, expectations for the distribution of HI at that redshift depend on our understanding of the cosmological HI mass density at the present epoch and its variation with environment.  We exploit a complete optical diameter and HI flux limited sample of galaxies in the local universe to derive a robust measurement of the HIMF for masses log($M / M_{\odot}$) $> 7.4$ which takes into account the effects of local large scale structure.
\end{abstract}
\thispagestyle{plain}

\section{Introduction}

This paper is an abbreviated discussion of the digital HI catalog presented in detail by Springob, Haynes, \& Giovanelli (2004a), and the HI mass function (HIMF) results presented in Springob, Haynes, \& Giovanelli (2004b).

\section{HI Sample}

Haynes et al. (1999) presented HI parameters for a set of 1201 galaxies contained in a 
digital archive of HI spectra maintained by Martha Haynes and Riccardo Giovanelli.  The spectra were taken at 
the Arecibo 305 meter Telescope, the Green Bank 91 meter Telescope, the Green Bank 43 
meter Telescope, the Effelsberg 100 meter Telescope, and the Nancay Radiotelescope.  We have produced an updated version of this archive, including observations made between 1981 and 2001.  The archive contains spectral parameters extracted from 21 cm line spectra for 8852 galaxies in the local universe (heliocentric velocity $-200 < V_{\odot} < 28,000$
km/s).  As in Haynes et al. (1999), we augment the sample of galaxies in the archive with HI fluxes from the literature as compiled in the all-sky private database that is also maintained by Haynes and Giovanelli known as the Arecibo General Catalog (AGC).  The total number of HI flux measurements available in this sample is 13,773.

In order to minimize the effecs of sample selection and bias, we select a sample of 2771 galaxies (hereafter referred to as the ``complete sample'') from these data, containing all objects in the HI digital archive or the AGC with $a > 1.0$ (where $a$ is the optical apparent diameter in arcminutes), $-2^{\circ} <$ decl. $< +38^{\circ}$, galactic latitude $|b| > 15^{\circ}$, morphological type Sa-Irr, and for which log$(F_{HI})>0.6$ (where $F_{HI}$ is the corrected HI flux density integrated over the profile in units of Jy km/s).  Ellipticals, S0s, and S0as are excluded due to the poor completeness of the archive for these types.

\section{Methods}

The $\Sigma (1/V_{max})$ method, first developed by Schmidt (1968), has frequently been applied to the determination of HI mass functions (e.g., Zwaan et al. 1997, hereafter Z97; Rosenberg \& Schneider 2002, hereafter RS).  We use a modified version of this method that uses the PSCz density reconstruction (presented in Branchini et al. 1999) to correct for the effects of large scale structure.  This approach is discussed in Masters, Haynes, \& Giovanelli (2004).

An alternative method for deriving the HIMF is the two-dimensional stepwise maximum likelihood (2DSWML) method, developed by Loveday (2000) for constructing an optical luminosity function, and first used for the HIMF by Zwaan et al. (2003, hereafter Z03).  Z03 used this method rather than $\Sigma (1/V_{max})$ because of the latter's insensitivity to the effects of large scale structure.  We believe this concern has been abrogated by our use of a modified version of $\Sigma (1/V_{max})$ that accounts for large structure, but we nonetheless have used the 2DSWML method as a check on our results.

\section {Results}

Figure 1 shows the HIMF, as determined by both the $\Sigma (1/V_{max})$ and 2DSWML methods, for the complete sample.  The normalization for the 2DSWML points has been chosen to minimize scatter with the $\Sigma (1/V_{max})$ points.  Agreement between the two methods appears to be quite good, with the 2DSWML points falling within 1$\sigma$ of the $\Sigma (1/V_{max})$ for 12 of the 15 bins.  We have fit the $\Sigma (1/V_{max})$ HIMF to a Schechter function (Schechter 1976) defined by

\begin{equation}
\phi(M_{HI}) = {dN \over d \; {\rm log}(M_{HI})} = {\rm ln}(10) \phi_{*} (M_{HI}/M_{*})^{\alpha + 1} {\rm exp} (-M_{HI}/M_{*})
\end{equation}

\noindent Our best fit Schechter parameters are $\alpha = -1.24$, log$(M_{*}/M_{\odot})=9.99$, $\phi_{*}=3.2 \times 10^{-3}$ Mpc$^{-3}$, with a $\chi_\nu^2$ of 0.99.



\begin{figure}[ht]
\plotfiddle{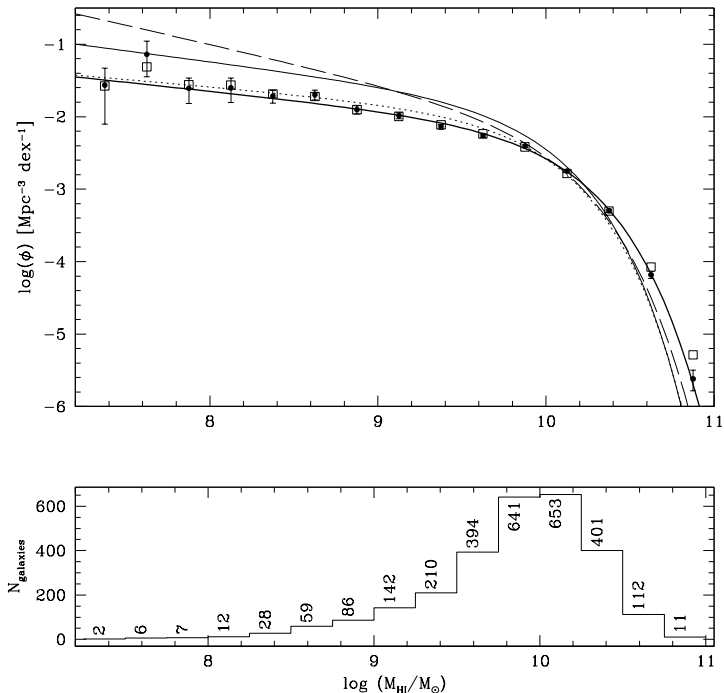}{3.0in}{0}{50}{50}{-160}{-125}
\vskip 1cm

\caption{{\it Top}, HIMF for the complete sample, using the $\Sigma (1/V_{max})$ method (filled circles) and the 2DSWML method (open squares).  $\Sigma (1/V_{max})$ errorbars indicate 1$\sigma$ uncertainties based on Poisson counting statistics.  The thick solid line is the best-fit Schechter function to the $\Sigma (1/V_{max})$ points.  Other lines show the Schechter parameters from three HI blind surveys.  (Z97: {\it dotted}; RS: {\it dashed}; Z03: {\it thin solid}); {\it bottom}, distribution of HI masses in the full sample per 0.25 dex bin.}

\end{figure}

In Figure 1 we have also overplotted the HIMFs of three HI selected samples: Z97, RS, and Z03.  At masses of $\sim 10^8 M_{\odot}$, both our HIMF and that of Z97 are lower than each of the others by a factor of a few.  Another way to compare the results is by comparing the total HI mass density, derived by integrating the Schechter function over all HI masses.  Using $\rho_{HI}=\phi_{*}\Gamma(2+\alpha)M_{*}$, we get $\rho_{HI}=3.8 \times 10^7 M_{\odot}$/Mpc$^{-3}$.  Converting this to $\Omega_{HI}$ gives $\Omega_{HI} = 2.7\times 10^{-4}$, 39\% lower than the value reported by Z03, which is based on the HI Parkes All Sky Survey (HIPASS) Bright Galaxy Catalog (BGC), the largest HI blind survey to date.

There are several possible explanations for this discrepancy.  It may be that galaxies with small physical diameters, not included in our complete sample because of our diameter limit, are large contributors to the low mass end of the HIMF.  In Springob, Haynes, \& Giovanelli (2004a) we show that, across most of the $M_{HI}$ distribution, the BGC includes some galaxies with smaller diameters than those included in our complete sample.  However, this is least apparent for log$(M_{HI}/M_{\odot}) < 8.5$, where the divergence between the HIMFs is greatest.

We also note that Briggs \& Rao (1993),  the only other work to estimate the HIMF based on a large optically selected sample including the range of morphological types we include here, finds a low mass slope of $\alpha \sim -1.25$, nearly identical to our value.  This suggests that the discrepancy may be due to some hidden systematic bias in the selection of optical samples.  HI selected samples may be detecting faint low surface brightness galaxies that are missed by optical surveys.  In particular, there may be galaxies that meet our diameter limit, but are not included in our sample because they are too faint to be optically detected.  Yet of the 758 BGC galaxies with $|b| > 15^{\circ}$, only 39 do not have optical counterparts with diameter estimates listed in the AGC.  And only 3 of those have log$(M_{HI}/M_{\odot}) < 8.5$.

One final possible explanation for the disagreement between our estimate of the low mass end of the HIMF and those of most HI selected samples is that the discrepancy is a consequence of the fact that the lowest mass galaxies in each of these samples are confined to a very local volume.  In both the BGC and our complete sample, all galaxies with log$(M_{HI}/M_{\odot}) < 8.5$ are found within $v_{helio} < 1200$ km/s.  There could be systematic distance errors due to defects in the flow model, or there could be systematic environmental effects within that local volume.  We note that the only HIMF derived from an HI blind survey which matches our result at low masses is Z97, which, like our sample and unlike HIPASS, includes nearby galaxies in the direction of Virgo.  Given both the distance uncertainties and our poor understanding the environmental dependence of the HIMF (discussed in Section 3.1), we cannot resolve the question of why our HIMF diverges from that of most HI blind surveys at the low mass end.  These effects can only be disentangled with the aid of a deeper sample that includes low mass galaxies at larger distances.

\subsection {Environmental Dependence}

We have also computed the HIMF for three subsets of the complete sample, divided by local matter density as determined by PSCz ($n<1.5$, $1.5<n<3.0$, and $n>3.0$), where $n$ is the local matter density normalized such that $n = 1$ represents the average density of the universe.  The resulting HIMFs are shown in Figure 2.  Best-fit Schechter parameters can be found in Table 1.  We find that the lowest density sample has a steeper low-mass slope and a higher value of $M_{*}$ than the two higher density samples.  This trend is consistent with the results of  Briggs \& Rao (1993), RS, and Davies et al. (2004), all of whom found evidence for a flatter HIMF in Virgo than in the field.  However, the uncertainties are large enough that this is a statistically marginal result.

In Springob, Haynes, \& Giovanelli (2004b), we argue that the flattening of the HIMF in the two higher density regimes is unlikely to be caused by morphological segregation, and that {\it if} it is a result of increased HI deficiency relative to the low density regime, then this would be an indication of increasing deficiency towards smaller linear optical diameters at the low mass end of the HIMF, a result consistent with HI deficiency in clusters being caused by galaxy-ICM interactions.  We cannot, however, rule out the possibility that the environmental dependence is caused by variation within the morphological subsamples independent of the gas stripping in high density regions commonly associated with deficiency.

\begin{table}[|ht]
\caption{Schechter Parameters}
\smallskip
\begin{center}
{\small
\begin{tabular}{lccccc}
\tableline
\noalign{\smallskip}
subsample   & $N_{galaxies}$     & $\alpha$   & log$(M_{*}/M_{\odot})$ & $\phi_{*} [$Mpc$^{-3}]$    & ${\chi_\nu}^2$ \\
\noalign{\smallskip}

\tableline
\noalign{\smallskip}

all  & 2771  &  -1.24 & 9.99 &  $3.2 \times 10^{-3}$ &  0.99 \\
$n<1.5$  & 1036  &   -1.38 & 10.07 &  $8.7 \times 10^{-4}$ &  0.35 \\
$1.5<n<3.0$  & 918  &   -1.13 & 9.92 &  $8.2 \times 10^{-3}$ &  0.71 \\
$n>3.0$  & 751  &   -1.24 & 9.95 &  $2.1 \times 10^{-2}$ &  0.65 \\
\noalign{\smallskip}

\tableline

\end{tabular}
}
\end{center}
\end{table}



\begin{figure}[ht]
\plotfiddle{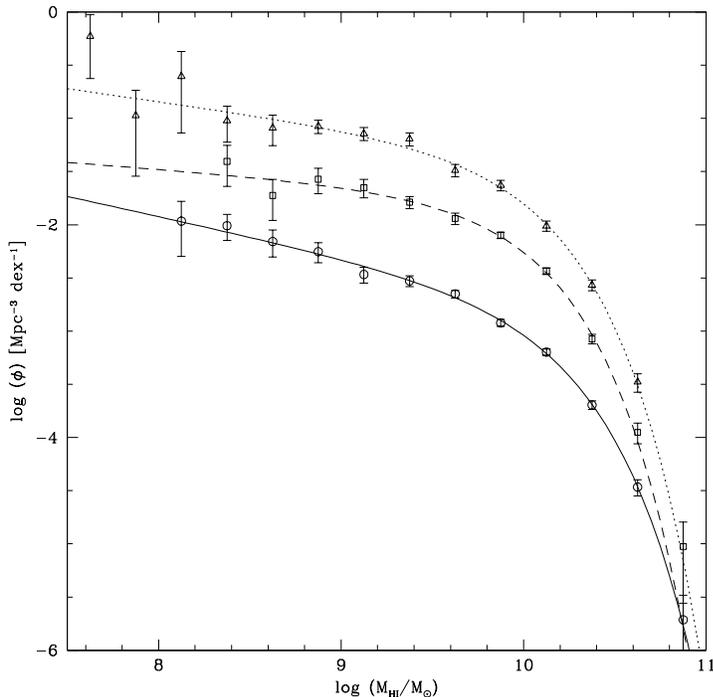}{3.0in}{0}{50}{50}{-160}{-125}
\vskip 1cm

\caption{HIMFs for the $n<1.5$ (open circles), $1.5<n<3.0$ (open squares), and $n>3.0$ (open triangles) samples.  Solid, long-dashed, and short-dashed lines show best-fit Schechter functions for each of the three samples respectively.}

\end{figure}

\section{Future Work}

The best way to resolve these open questions about the HIMF is with better statistics.  In particular, better statistics for galaxies at the low mass end of the HIMF would aid in resolving the question of why we detect fewer objects at low masses than most HI blind surveys, and in quantifying the environmental dependence of the HIMF.  The Arecibo Legacy Fast ALFA Survey, to be conducted from 2005-2010 with the Arecibo L-band Feed Array (ALFA)
will sample low mass galaxies more deeply in redshift than all previous surveys, and should allow us to answer these unresolved questions.

\vskip 0.2in

We wish to thank Enzo Branchini for providing a copy of the PSCz density reconstruction.  This work is partially funded by NSF grants AST-0307396 and AST-0307661, and the NRAO/GBT03B-007 Graduate Student Support Grant.

\vfill\pagebreak

\end{document}